\documentstyle[12pt]{article}

\newlength{\dinwidth}
\newlength{\dinmargin}
\begin{document}
\newcommand{\ra}{\rightarrow}
\newcommand{\tb}{tan \beta}
\newcommand{\mr}{{\stackrel{<}{\sim}}}
\begin{flushright}
IFT 15/96\\[1.5ex]
{\large \bf hep-ph/9607268} \\
\end{flushright}

\baselineskip 0.4cm
~\\
\vspace{2cm}
\begin{center}
  \begin{large}
  TWO HIGGS DOUBLET MODELS{\footnote{Invited talk at the
XXXIst RENCONTRES DE MORIOND, Electroweak Interactions and Unified Theories,  
Les Arcs, France, March 16-23, 1996}} \\
\end{large}
~\\
~\\
 {Maria Krawczyk}\\
Institute of Theoretical Physics\\
University of Warsaw, Poland\\
\vspace{2cm}

\vskip 2cm
{ABSTRACT} 
\end{center} 

\noindent
Present data do not rule out the light neutral Higgs particle
$h$ or $A$ in the framework of the general 2HDM. 
The discovery reach/exclusion limits 
of the Yukawa process $Z\rightarrow f {\bar f} h/A$,
($f= b$ quark or $\tau$ lepton) at LEP I and
of the gluon-gluon fusion at HERA is discussed.
In addition  the possible search for very light Higgs particle
in $\gamma \gamma$ fusion at low energy NLC  is described.

\newpage

\topmargin -60pt
\headheight 0pt



\baselineskip 0.6cm
\noindent
{1. INTRODUCTION}\\
The mechanism of spontaneous symmetry breaking  proposed as
the source of mass for the gauge and fermion fields in the Standard 
Model (SM) leads to  a neutral scalar particle,
the minimal Higgs boson.  According to  the LEP I data,
based on the Bjorken process $e^+e^- \ra H Z^*$,
it   should be heavier than $62.5$ GeV$^{1)}$.

The  minimal extension of the Standard Model is to include
a second Higgs doublet to the symmetry breaking
mechanism. In two Higgs doublet models 
the observed Higgs sector is enlarged to five scalars: two
neutral Higgs scalars (with masses $M_H$ and $M_h$ for heavier and
lighter particle, respectively), one neutral pseudoscalar
($M_A$), and a pair of charged Higgs's ($M_{H^+}$ and $M_{H^-}$). 
 The
neutral Higgs scalar couplings to quarks, charged leptons and gauge 
bosons are 
modified with respect to analogous couplings in SM by factors that 
depend on additional parameters : $\tan\beta$, which is
the ratio of the vacuum expectation values of the Higgs doublets
 $v_2/v_1$,
and the mixing angle in the neutral Higgs sector $\alpha$. Further,
 new couplings appear : Z$h (H) A$ and Z$H^+ H^-$.

In this talk I will focus on the  appealing version of the models
with two doublets ("type II")  with one Higgs doublet
with vacuum expectation value $v_2$ couples only to the "up"
components 
of fermion doublets while the other one couples to the "down" 
components. 
The large top quark mass can  easily  beaccommodated in such scenerio for  
large ratio $v_2/v_1 \sim m_{top}/m_b\gg 1$.
The well known supersymmetric model (MSSM) belongs to this  class. 
In this model the relations among the parameters required by the 
supersymmetry appear, leaving only two parameters free
e.g. M$_h$ and tan$\beta $.
In general case, which we call the general 2 Higgs Doublet Model
 (2HDM), masses and parameters $\alpha$ and $\beta$ 
are not constrained by the model.
Therefore the same experimental data may lead to very distinct 
consequences  depending on the
considered versions of two Higgs doublet extension of SM:
supersymmetric and nonsupersymmetric one.

The current mass limit on {\underline {charged}} Higgs boson $M_{H^{\pm}}$
44 GeV/c was obtained$^{2)}$
from process $Z \ra H^+H^-$, 
which is independent on the additional parametres $\alpha$ and 
$\beta$. 
(Note that in
the MSSM version one expect 
$M_{H^{\pm}} > M_W$). For {\underline {neutral}} Higgs particles $h$ and 
$A$ there are two
main sources  of information:  
 the Bjorken processes Z$ \ra Z^*h $
which constraints  $g_{hZZ}^2 \sim sin^2(\alpha-\beta)$
(which was found to be
 smaller than 0.1 for the 0$\mr  M_h\mr $50 GeV)
and the  process Z $\ra hA$
constraining the $g_{ZhA}^2 \sim cos^2(\alpha-\beta)$$^{3)}$.
These results  can be translated into
the limits on neutral Higgs bosons masses
 M$_h$ and M$_A$.
 
 In the MSSM, due to relations among parameters, 
the above data allows to draw limits for the masses
 of individual particles: $M_h\ge 44$ GeV/c for any tan $\beta $ 
 and $M_A \ge$ 39 GeV/c for tan$\beta \ge$1 $^{4)}$.
In the general 2HDM the implications are quite different. 
The large portion of the (M$_h$,M$_A$) plane,
where both masses are in the range between 
0 and $\sim$50 GeV, is excluded as well$^{2)}$.
However in this case, with Higgs scalar production via Bjorken 
process suppressed  by the small factor $sin^2(\alpha -\beta)\sim 0$
and the neutral Higgs bosons pair production  forbiden kinematically,
there are not any experimental limits from LEP I 
on  {\underline {individual}}  masses $M_h$ and $M_A$.
So $M_h$  
can be arbitrarily small provided $M_A$ is sufficiently heavy
(i.e. $M_h+M_A>M_Z$), or vice versa $^{5-6)}$.

{\it Summarizing, the precise LEP I data still leave the window for one 
light Higgs 
particle (neutral scalar $h$ or pseudoscalar $A$) 
which is open {\underline {only}} for the 2HDM.}

\noindent
{2. THE 2HDM  WITH LIGHT HIGGS PARTICLE}\\
We  consider the possibility of the existence of a
light neutral Higgs particle with mass below $\sim$ 50 GeV,
and  specify the model further  by choosing particular 
values for the parameters $\alpha$ and $\beta$
within the present limits.
We  take for simplicity $\alpha=\beta$.
This assumption  leads to equal in  
 strengths of the coupling of  fermions to scalars   and pseudoscalars.
{{In 2DHM fermions couple to the pseudoscalar 
with a strength  proportional to $(tan \beta)^{\pm1}$
whereas the coupling of the fermions to the scalar $h$
goes as $(sin \alpha/cos \beta)^{\pm1}$, where the sign
$\pm$  corresponds to  isospin $\pm$1/2 components}}. 
We will consider the scenario with 
large $tan\beta \sim {\cal O}(m_t/m_b)$ and therefore with  large
 enhancement 
in the coupling of both $h$ and $A$ bosons to the down-type
 quarks and leptons.
The Higgs particles decay
mainly into $\tau^+\tau^-$ for $4<M_{h/A}<10$ GeV and into $b {\bar b}$ 
above 10 GeV
with branching ratios  close to 1$^{5)}$. 
\begin{figure}[ht]
\vskip 4.23in\relax\noindent\hskip -1.25in
	     \relax{\includegraphics{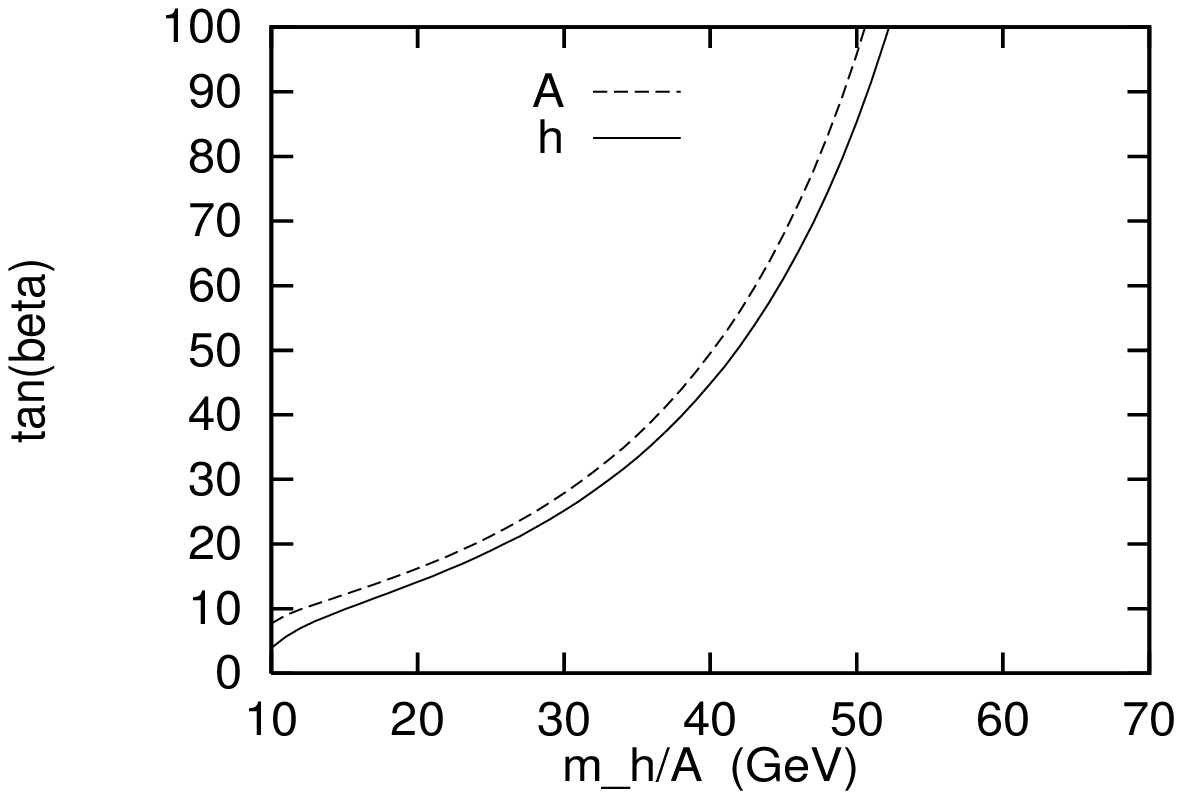}}
	     \relax\noindent\hskip 3.15in
	     \relax{\includegraphics{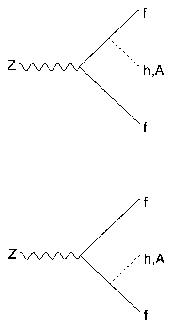}}
\vspace{-17.5ex}
\baselineskip 0.4cm
\caption{ {\em The 95\% C.L. exclusion plot for $h$ and $A$, based on Ref.12, 
the part above curves is excluded
} }
 \label{fig:signal}
\baselineskip 0.6cm   
\end{figure}

\noindent   
{3. CONSTRAING THE MODEL }\\  
The existing 
limits for large $\tan \beta$ and low mass Higgs scalar/pseudoscalar 
scenario in 2HDM are very weak, below I comment shortly some
low energy data. The discussion of the potential 
of future measurements follows then.\\
\noindent   
{\underline {\it Low energy data}}\\
** {New data on B$\ra \tau \nu$ X gives tan$\beta\leq
 0.52$ GeV$^{-1} M_{H^{\pm}}$$^{7)}$. Thus 
  even  with  the lower limit of M$_{H^\pm}\sim 44$ GeV,
a large value of tan $\beta$ $\sim$20 is still allowed.}\\
** $g-2$ data for muon
can accomodate 
{large value of tan$\beta$ (20 or more)  for the Higgs boson masses 
equal or larger then 2 GeV
(see e.g. the discussion in Refs.13b and 8).\\  
**{The theoretical uncertainty of QCD and relativistic 
 corrections does not allow to use a so called Wilczek process 
 $\Upsilon \ra \gamma h(A)$$^{10)}$ to set decisive limits on 
 low mass  Higgs particle coupling to $b$ quark}.\\
\noindent
{\underline {\it { Yukawa process at LEP I}}\\
As we mentioned  LEP I data
set already important limits on the parameters of MSSM
as well as  2HDM$^{2-4)}$.
Still the decisive search as far as light Higgs particle
is concerned has to be performed. The reason is that 
in  the scenario we consider, at LEP I
the dominant production   of the neutral Higgs
boson $h(A)$ for the $M_h+M_A>M_Z$
is   a  bremsstrahlung from the {\it b} quark,
$e^+e^- \rightarrow b {\bar b} h(A)$, or the analogous one 
 with a bremsstrahlung from $\tau$ lepton.
{{Note that this process is being now analized by experimental groups
L3 and Aleph$^{11)}$}. The
importance of this process, which we will called {\it Yukawa process}, for  
the search of light non-minimal neutral Higgs particle 
 at LEP I was stressed in $^{5)}$ (see also Ref.9).
The potential of this process is discussed in details in Ref.12. Fig.1 
presents the exclusion plot(95\%C.L.)  obtained using $\tau \tau bb$ final 
state with the QCD corrections and fermion mass terms included.\\ 
\noindent
{\underline {\it gluon-gluon fusion at HERA}}\\
The gluon-gluon fusion
  via a quark loop,
$gg \ra h(A)$,
can be  a significant source of light non-minimal neutral Higgs bosons
at HERA  collider due to the hadronic
interaction of quasi-real photons with protons$^{13)}$.
\begin{figure}[ht]
\vskip 4.45in\relax\noindent\hskip -0.98in
	     \relax{\includegraphics{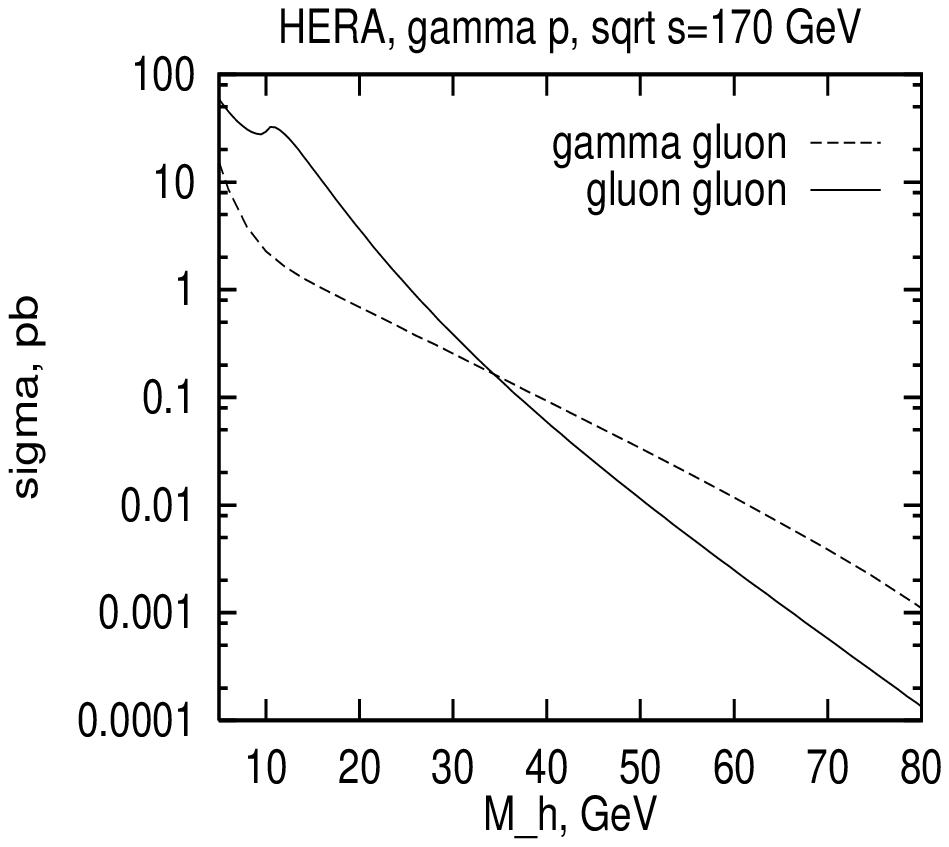}}
	     \relax\noindent\hskip 2.35in
	     \relax{\includegraphics{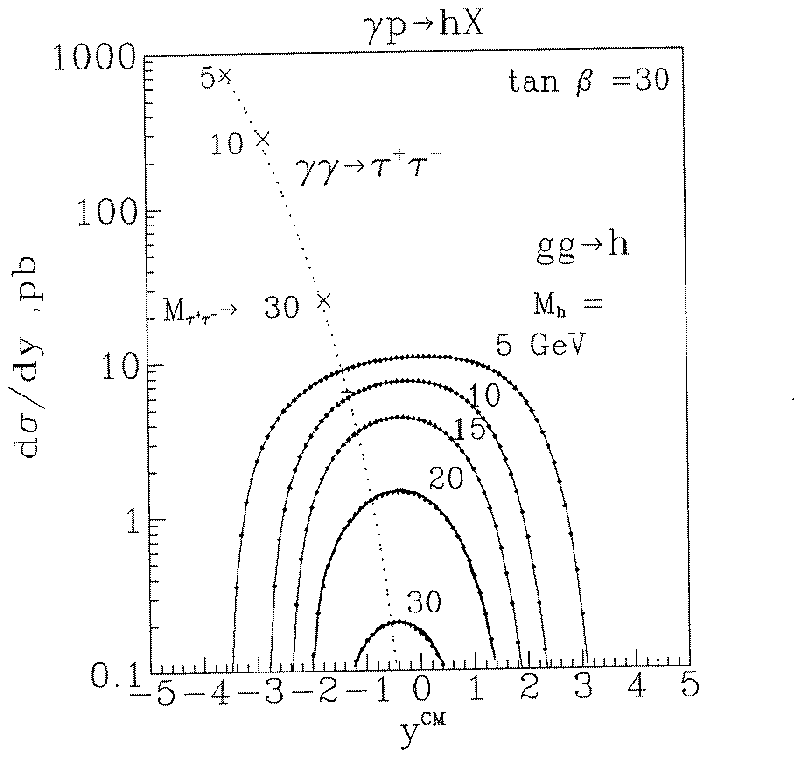}}
\vspace{-17.5ex}
\baselineskip 0.4cm
\caption{ {\em The  cross sections for the Higgs scalar production
 ($\tan \beta = 30$) a) total, b) rapidity distribution.
 The GRV parametrizations are assumed for the photon$^{14a)}$ 
 and for the proton$^{14b)}$. 
 } }
    \label{fig:sig}
\end{figure}
\baselineskip 0.6 cm
In addition the  production of the neutral Higgs boson via
$\gamma g \ra b {\bar b} h $
may also be substancial$^{6,13)}$. Note that the latter process
also includes  
 the lowest order contributions due to the resolved photon,
like $\gamma b\ra bh$, $b {\bar b}\ra h$, $bg \ra hb$
etc. 
We study  both $gg$ and $\gamma g$ fusions in a tagged mode of HERA.
In Fig.2a the total cross
section for $\gamma p \ra h X$  at $\sqrt{S_{\gamma p}}$=170 GeV
is plotted for tan $\beta$=30.
Note that for mass below $\sim 30$  GeV the $gg$
clearly dominates.
In order to detect the Higgs particle 
it is useful to
study the rapidity distribution ${d \sigma }/{dy}$ of the Higgs bosons
in the $\gamma p$ centre of mass system.
Note that $y=-{{1}\over{2}}log{{E_h-p_h}\over{E_h+p_h}}
=-{{1}\over{2}}log{{x_{\gamma}}\over{x_p}}$, where $x_p(x_{\gamma})$
are the ratio of energy of gluon to the energy of the proton(photon),
respectively.
The (almost)
symmetric shape of  the rapidity distribution found for the signal
is extremely useful
 to reduce the
background and to separate the $gg\rightarrow h$ contribution$^{13)}$(Fig.2b).
These results hold for the $A$ production as well.

The main background for Higgs mass range between
$\tau$ and $b$ thresholds is due to $\gamma\gamma\rightarrow \tau^+\tau^-$,
with the $\gamma$ flux from the proton (given by the
elastic equivalent photon  approach of reference $^{15a)}$ or 
{{ by the standard Weizs\"{a}cker-Williams
  spectrum $^{15b)}$}}).
In the region of negative rapidity 
the cross section is very large $\sim$ 800 pb
at the edge of phase space, then it falls down rapidly
approaching $y=0$. At the same time signal reaches at most 10 pb
(for $M_h$=5 GeV).
The region of positive rapidity  is {\underline {not}} allowed 
kinematically for this process since here one photon interacts directly with 
$x_{\gamma}=1$, and therefore $y_{\tau^+ \tau^-}
 =-{{1}\over{2}}log{{1}\over{x_p}}\leq 0$.
 Moreover, there is a simple
 relation between rapidity and invariant mass:
 $M^2_{\tau^+ \tau^-}=e^{2y_{\tau^+ \tau^-}}S_{\gamma p}$.
Significantly different topology found for these events
than for the signal  
 allow to get rid of
 the background due to process $\gamma\gamma
\rightarrow \tau^+\tau^-$. 
The other sources of background are 
$q\bar q\rightarrow\tau^+\tau^-$ processes.
These processes contribute to
 positive and negative
rapidity y$_{\tau^+ \tau^-}$, with a flat and
relatively low cross sections in the central region.

The results for the tagged
case indicate that for
$\tan\beta\simeq30$ there is a chance to observe the Higgs boson at 
HERA for $M_h {\stackrel{<}{\sim}} 30$ GeV.
Assuming that the luminosity ${\cal L}_{\gamma p}$=20 pb$^{-1}$/y
 we predict that $gg$ fusion 
will produce approximately 900 events
per annum for $M_h=5$ GeV (10 events  for
$M_h=30$ GeV).
A clear signature for the $\tau^+\tau^-$ final state 
 at positive centre-of-mass
rapidities of the Higgs particle should be seen, even for the mass 
of Higgs particle above the $bb$ threshold
(more details can be found in Ref.13).\\
\begin{figure}[ht]
\vskip 4.45in\relax\noindent\hskip -1.55in
	     \relax{\includegraphics{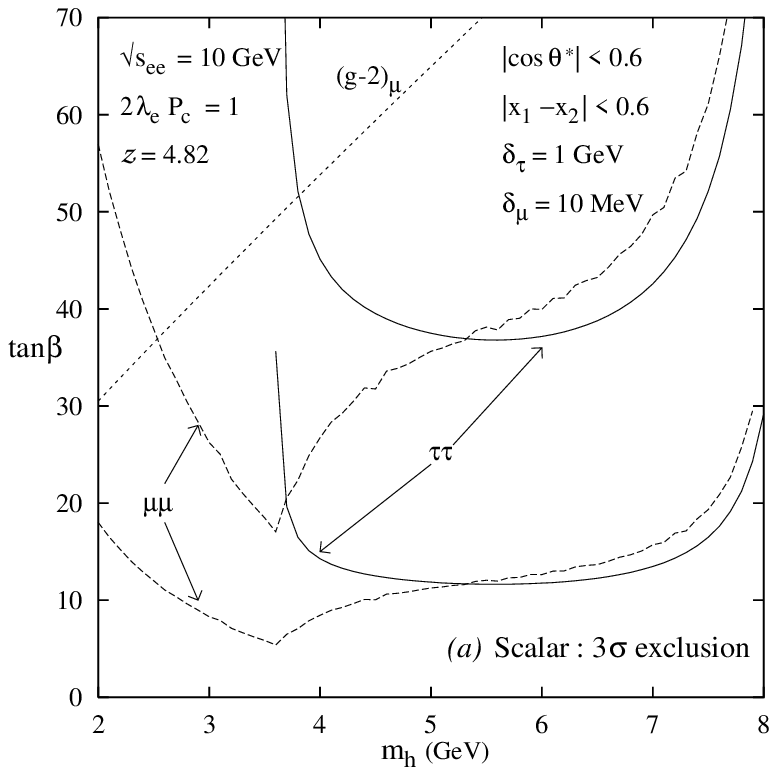}}
	     \relax\noindent\hskip 3.35in
	     \relax{\includegraphics{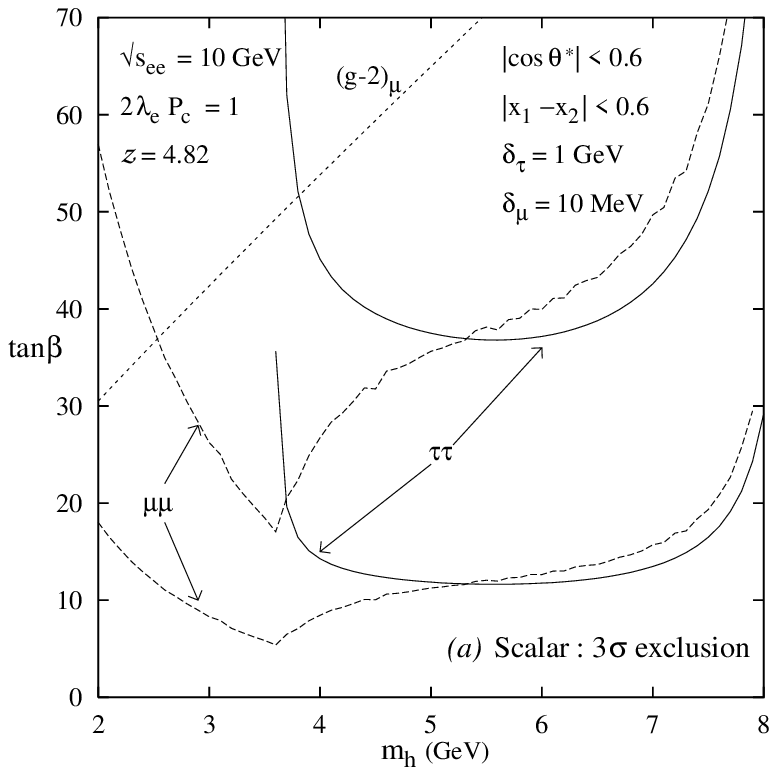}}
\vspace{-22.5ex}
\baselineskip 0.4cm
\caption{
{\em The exclusion plots for  {\em (a)} scalar 
and {\em (b)} pseudoscalar for $\mu \mu$-- and $\tau \tau$--channels.
Parameter space above the curves can be ruled out at $3 \sigma$. The upper
and lower sets are for integrated luminosity of 100 pb$^{-1}$ and
10 fb$^{-1}$ respectively. The limits { derivable} from present 
$(g - 2)_\mu$ measurement are also shown for comparison.
results from Ref.16} }
\label{fig:excl}
\end{figure}
\baselineskip 0.6cm
\noindent
{\underline {\it photon-photon fusion at NLC}}\\
The posible search for a {\it very} light Higgs particle may
in principle be performed at low energy option of NLC.
In the paper $^{16)}$ we address this problem
and find that the exclusion based on the $\gamma \gamma$ fusion
into Higgs particle decaying into $\mu \mu$ pair may be very efficient 
in probing the value of tan $\beta$  down to 5 at $M_h\sim 3.5$ GeV
and below 15 for $2.5<M_h<7.5$ GeV (3 $\sigma$ exclusion)
provided that the luminosity is equal to 10 fb$^{-1}$/y (see Fig.3). 

\noindent
{3.CONCLUSION}\\
  Since in framework of 2HDM
   even very light neutral Higgs particle
  is not rule out by present data 
  the search for it should be performed.
  The most suitable, at least for mass larger than 10 GeV,
  seems to be the Yukawa process at LEP I.
  The other opprortunity may be HERA collider, where
  the production occurs due to the structure of photon in $gg$ fusion
  and in $\gamma g$ fusion. For very light masses the 
  low energy $\gamma \gamma$ NLC
maight be used.

\noindent
{ACKNOWLEDGMENT}\\
The results have been obtained
in collaboration with Ahmed C. Bawa, Jan Kalinowski,  Debajyoti Choudhury
and Jan \.Zochowski. I am especially grateful to Jan Kalinowski for 
many fruitful discussions.
I would like to thank organizers of the Rencontres de Moriond for 
their kind invitation and financial support.

\noindent   
REFERENCES\\
\baselineskip 0.4cm

\noindent
1. Brussels Conference, June 1995, J. -M. Grivaz talk\\ 
2. A. Sopczak, Int. Workshop "Physics from Planck Scale to
 Electro-Weak Scale",
Warsaw, Poland, Sept. 1994,DELPHI CERN-PPE/94-83\\
3. ALEPH Coll., D. Buskulic et al., Z. Phys. C 62 (1994) 
539;
DELPHI Collab., P. Abreu et al., Nucl. Phys. B 418 (1994) 403; L3 
Collab.,
 M. Acciarri et al., Z. Phys. C 62 (1994) 551; Note 1800 (1995)
OPAL Collab. R. Akers et al., Z. Phys. C 61 (1994) 19\\
4. Aleph Coll. Note EPS0415(1995);L3 Coll. Note 1801(1995)\\
5. J. Kalinowski and S. Pokorski, Phys. Lett. {\bf B219}
 (1989) 116;
J. Kalinowski, N. P. Nilles, Phys. Lett. B 255 (1991) 134;
A. Djouadi et al,  Phys. Lett.
 {\bf B259} (1990) 175.\\
6.B. Grz\c{a}dkowski, S. Pokorski and J. Rosiek, Phys. Lett.
 {\bf B272} (1991)174\\
7. Aleph Coll. CERN-PPE/94-165\\
8. U.Chattopadhyay and P.Nath, Phys. Rev. D53 (1996) 1648\\
9. E. Carlson, S. Glashow, U. Sarid, Nucl.Phys. B 309 (1988)597\\
10.  D. Antresyan et al., Phys. Lett.
B251(1990)204;  R. Balest et al., CLNS 94/1292, CLEO 94-19 \\
11. L3 Coll. Note 1801(1995); A. Sopczak, J. M. Grivaz, 
private communication\\
12. J. Kalinowski, M. Krawczyk, Phys.Lett B361 (1995)66;
Acta Phys.Pol. B27 (1996) 961\\
13. A.C. Bawa and M. Krawczyk, Warsaw IFT preprints 
16/91 and ERRATUM, 
16/92; Phys. Lett B357 (1995) 637; M. Krawczyk, talk in 
Kazimierz Symposium'95 (Ames, USA, May 95).\\
14a.  
M. Glueck, E. Reya, A. Vogt , Phys. Rev. D46 (1992)
 1973;
14b. Phys.Rev.D46(1992)1973;\\
15a. B.A. Kniehl, Phys. Lett. {\bf B254} (1991) 267;
15b. E. J. Williams, Proc. Roy. Soc. London (A) 139 (1933) 163; 
Phys. Rev.45 (1934) 729(L); K. Dan. Vidensk.Selsk.Math.-fys. Medd.
13, No. 4 (1935);
C. F. v. Weizs\"acker, Z. Phys. 88 (1934) 612\\
16. D. Choudhury, M. Krawczyk, MPI-Pth/96-46; Proc. Workshop
NLC, 1995, p.69; and in preparation
\end{document}